\documentclass[aps,prl,reprint,groupedaddress,showpacs]{revtex4-1}

\usepackage{dcolumn}
\usepackage{amssymb,url}
\usepackage[dvipdfmx]{graphicx}
\usepackage{bm}

\newcommand{\bra}{\langle}
\newcommand{\ket}{\rangle}
\begin{document}

%Title of Rapid Communication
\title{Thermal Pure Quantum States of Many-Particle Systems}


\author{Masahiko Hyuga}
\email[]{hyuga@ASone.c.u-tokyo.ac.jp}
\author{Sho Sugiura}
\email{sugiura@ASone.c.u-tokyo.ac.jp}
\author{Kazumitsu Sakai}
\email{sakai@gokutan.c.u-tokyo.ac.jp}
%\thanks{}
\author{Akira Shimizu}
\email{shmz@ASone.c.u-tokyo.ac.jp}
\homepage[]{http://as2.c.u-tokyo.ac.jp}
\affiliation{Department of Basic Science, 
University of Tokyo, 3-8-1 Komaba, Meguro, 
Tokyo 153-8902, Japan}


\date{\today}

\begin{abstract}
We generalize the thermal pure quantum (TPQ) formulation 
of statistical mechanics,
in such a way that it is applicable to systems 
whose Hilbert space is infinite dimensional.
Assuming particle systems, 
we construct the grand-canonical TPQ (gTPQ) state,
which is the counterpart of the grand-canonical Gibbs state of the ensemble formulation.
A single realization of the gTPQ state gives 
all quantities of statistical-mechanical interest,
with exponentially small probability of error. 
This formulation not only sheds new light on quantum statistical mechanics
but also is useful for practical computations.
As an illustration, 
we apply it to the Hubbard model,
on a one-dimensional ($1d$) chain and on a 
two-dimensional ($2d$) triangular lattice.
For the $1d$ chain, our results 
agree well with the exact solutions 
over wide ranges of temperature, chemical potential 
and the on-site interaction. 
For the $2d$ triangular lattice, 
for which exact results are unknown, 
we obtain reliable results over a wide range of temperature.
We also find that finite-size effects are much smaller in
the gTPQ state than in the canonical TPQ (cTPQ) state.
This also shows that 
in the ensemble formulation 
the grand-canonical Gibbs state of a finite-size system 
simulates an infinite system much better than the canonical 
Gibbs state.
\end{abstract}

%\keywords{}

\maketitle

% INTRODUCTION
%
Quantum statistical mechanics has conventionally been 
formulated as the ensemble formulation, in which %. In that formulation, 
an equilibrium state is given by a mixed quantum state 
(Gibbs state) 
that is represented by a density operator $\hat{\rho}^{\rm ens}$. 
Recently, another formulation, called the TPQ formulation, 
has been developed by 
two of the authors \cite{SS2012,SS2013},
by generalizing theories of typicality 
\cite{Popescu,Lebowitz,SugitaJ,SugitaE,Reimann2007,Reimann2008}. 
In this formulation, an equilibrium state is given by a pure quantum state, 
which is called a TPQ state. 
Since the TPQ state is not a purification \cite{NC} of $\hat{\rho}^{\rm ens}$,
it is totally different from $\hat{\rho}^{\rm ens}$.
In fact, the magnitudes of their entanglement 
are almost maximally different \cite{SS2005, Kindai2013}.
Nevertheless, 
one can correctly obtain all quantities of statistical-mechanical interest,
including thermodynamic functions, from 
a single state vector of a TPQ state \cite{SS2012,SS2013}.
Because of this striking property, 
the TPQ formulation is very useful in practical applications \cite{SS2012,SS2013}. 
In fact, it has solved problems that are hard with conventional methods,
such as 
the specific heat of a $2d$ frustrated spin system \cite{SS2013}.

However, it was formulated only for systems 
whose Hilbert space $\mathcal{H}$ is finite dimensional.
Since $\dim \mathcal{H} = \infty$ for many physical systems, 
such as particles in continuous space, 
generalization of the TPQ formulation is necessary.
Furthermore, 
only the microcanonical TPQ (mTPQ) 
and cTPQ states were constructed, and their validity 
was confirmed separately \cite{SS2012,SS2013}. 
Although all TPQ states give the same results 
in the thermodynamic limit \cite{SS2013}, 
they will give different results for finite-size systems 
because of finite-size effects.
To study infinite systems, 
it is desirable to develop 
other TPQ states (such as the gTPQ state) 
and to clarify which TPQ state of finite size gives results
closest to those of infinite systems.

In this Rapid Communication, 
we generalize the TPQ formulation so that 
it will be applicable to the case where 
$\dim \mathcal{H}$ and the norm of operators (such as 
the momentum) are infinite.
Assuming particle systems as a concrete example,
we construct the gTPQ state,
which are specified by 
inverse temperature $\beta =1/T$, 
chemical potential $\mu$,
volume $V$, 
magnetic field $\bm{h}$, and so on.
[In the following, 
we abbreviate $\beta, \mu, V, \bm{h}, \ldots$
simply as $\beta, \mu, V$.]
We show that a single realization of the gTPQ state gives 
all quantities of statistical-mechanical interest, 
including thermodynamic functions.
This striking property is not only interesting as a fundamental physics,
but also 
useful for practical computations,
because it enables one to solve problems that are hardly
solvable by other methods.
As an illustration, 
we apply the TPQ formulation to numerical 
studies of the Hubbard model,
on a $1d$ chain and on a $2d$ triangular lattice.
% For the $1d$ chain, our results 
% for the particle density
% and the specific heat, 
% obtained with the gTPQ state, 
% agree well with the exact results, 
% over wide ranges of 
% $T, \mu$ and the on-site interaction $U$. 
% We also obtain the spin and charge correlation functions,
% for which exact solutions are unknown.
% For the $2d$ triangular lattice, 
% for which exact results are unknown, 
We obtain reliable results, % with sufficiently small errors,
over wide ranges of 
$T, \mu$ and the on-site interaction $U$. 
% over a wide range of $T$.
Moreover, we show that 
as compared with the cTPQ state with finite $V$
the gTPQ state with the same $V$ 
gives results much closer to the exact results for an infinite system.
The same can be said for 
the canonical and grand-canonical Gibbs states 
of the ensemble formulation.

{\em Mechanical variables -- }
Statistical mechanics treats 
`mechanical variables', such as energy, and 
`genuine thermodynamic variables', such as entropy. 
Unfortunately,
the general definition of mechanical variables
in the previous formulation \cite{SS2012,SS2013}
% For mechanical variables,
% their definition  
% in the previous works
% \cite{SS2012,SS2013,Popescu,Lebowitz,SugitaJ,SugitaE,Reimann2007,Reimann2008}
breaks down when 
$\| \hat{A} \| = \infty$.
Therefore, we here define them more physically as follows \cite{degreeA}.
Let $\hat{A}$ be a low-degree polynomial 
(i.e., its degree is $\Theta(1)$)
of local observables. 
[For the order symbols, see, e.g., Ref.~\cite{NC}.]
We make it dimensionless.
For example, 
we denote by $\hat{H}$ the original Hamiltonian divided by 
an appropriate energy (such as the transfer energy). % $E_{\rm unit}$.
We call $\hat{A}$ a {\em mechanical variable}
if there exist a function $K(\beta, \mu)$
and a constant $m$,
both being positive and independent of $\hat{A}$ and $V$,
such that
\begin{equation}
\langle{ \hat{A}^2 }\rangle^{\rm ens}_{\beta \mu V}
\leq 
K(\beta, \mu) V^{2m}
\mbox{ for all } \beta, \mu, V.
\label{bound_<A>}
\end{equation}
This means that in an equilibrium state $\hat{A}$ 
should have finite expectation value and fluctuation 
even if $\| \hat{A} \| = \infty$.
For example, 
$n$-point correlation functions with $n \leq \Theta(m \ln V)$ 
(such as the spin-spin correlation function), 
and their sum (such as $\hat{H}$), are mechanical variables.
% whereas foolish operators (such as $e^V \hat{H}$) are not.
% In contrast to Refs.~\cite{SS2012,SS2013},
% $\dim \mathcal{H}$
%% the dimension of the Hilbert space $\mathcal{H}$,
% and $\| \hat{A} \|$ can be infinite
% according to (\ref{bound_<A>}),
% which physically means that  in an equilibrium state every $\hat{A}$ 
% should have finite expectation value and fluctuation 
% even if $\| \hat{A} \| = \infty$.

{\em gTPQ state -- } 
% {\em Grand-canonical TPQ state -- } 
% We now construct a gTPQ state.
We consider many particles % fermions or bosons that are 
confined in 
a box of arbitrary spatial dimensions. 
We assume that the grand canonical Gibbs state 
$\hat{\rho}^{\rm ens}_{\beta \mu V}$ %ensemble formulation 
gives the correct results, 
which are consistent with thermodynamics \cite{TD}.
This implies, for example, that specific heat is positive.

Let $\{ | \nu \ket \}_\nu$ 
be an arbitrary orthonormal basis of $\mathcal{H}$.
Many equations (such as the main result Eq.~(7) of Ref.~\cite{SS2013})
of the previous formulation \cite{SS2012,SS2013}
% works 
% \cite{SS2012,SS2013,Popescu,Lebowitz,SugitaJ,SugitaE,Reimann2007,Reimann2008}
% the random vector 
% $\sum_\nu c_\nu | \nu \ket$ (with $\sum_\nu |c_\nu|^2 = 1$) that 
% was used to construct TPQ states in Refs.~\cite{SS2012,SS2013}
become ill defined and/or meaningless
when $\dim \mathcal{H}=\infty$.
%  (and thus $\| \hat{A} \| = \infty$ for some $\hat{A}$'s).
% because $\sum_\nu |c_\nu|^2 = \dim \mathcal{H}=\infty$.
To overcome this difficulty, 
we first cut off `far-from equilibrium parts' of $| \nu \ket$ 
% by applying % the operator 
% $\exp[ -\beta (\hat{H} - \mu \hat{N})/2]$,
% where $\hat{N}$ is the number operator,
as
\begin{equation}
| \nu; \beta, \mu, V \ket 
\equiv \exp[ -\beta (\hat{H} - \mu \hat{N})/2]
| \nu \ket,
\end{equation}
where $\hat{N}$ is the number operator.
We then superpose $| \nu; \beta, \mu, V \ket $'s as
\begin{equation}
|\beta \mu V \ket
\equiv
\sum_\nu z_\nu | \nu; \beta, \mu, V \ket .
% =
% \sum_\nu z_\nu \exp[ -\beta (\hat{H} - \mu \hat{N})/2] | \nu \ket.
\label{gTPQ}
\end{equation}
Here, $z_\nu \equiv (x_\nu + i y_\nu)/\sqrt{2}$, 
where
$x_1, x_2, \ldots$ and $y_1, y_2, \ldots$ are 
real random variables, 
each obeying the unit normal distribution.
% We now show that 
% this vector is well defined 
% and that it is a gTPQ state.
We first show that 
this vector is well defined, i.e., 
its norm
% The norm of this vector % $|\beta \mu V \ket$ 
is finite for finite $V$
even when $\dim \mathcal{H} = \infty$, % (such as boson systems),
with probability that approaches one with increasing $V$. 
% in the thermodynamic limit.
(By contrast, the norm of another random vector
$\sum_\nu z_\nu | \nu \ket$ diverges with $\dim \mathcal{H}$.)
To show this, 
we invoke a Markov-type inequality:
Let $x$ be a real random variable and $y$ a real number, 
then for arbitrary ${\epsilon}>0$,
\begin{equation}
{\rm P} \left( 	| x - y | \geq \epsilon \right)
\leq
\overline{(x-y)^2}/ \epsilon^2,
\label{Markov}
\end{equation}
where the overbar denotes the random average.
Taking 
$
x = 
\langle \beta \mu V | \beta \mu V \rangle
/
\Xi
$
and $y=1$,
where $\Xi(\beta, \mu, V)$ is the grand-partition function,
we evaluate 
$
B_V^2
\equiv
\overline{
(\langle \beta \mu V | \beta \mu V \rangle
/
\Xi
-1)^2
}
$
as
\begin{equation}
B_V^2
\leq
1/
\exp [2V \beta \{ j(T/2, \mu ;V)-j(T, \mu ;V) \}].
% \exp [2V \beta \{ j(1/2\beta, \mu ;V)-j(1/\beta, \mu ;V) \}].
\label{Var<>}
\end{equation}
Here, % $T = 1/\beta$ (we take $k_{\rm B}=1$),
% $\Xi(\beta, \mu, V)$ is the grand-partition function,
$j(T, \mu; V) \equiv - (T/V) \ln \Xi(\beta, \mu, V)$
% $j(T, \mu; V) \equiv - (T \ln \Xi)/V$
is a thermodynamic function,
which 
%  ($T = 1/\beta$; we take $k_{\rm B}=1$).
% [The argument $(T, \mu; V)$ indicates that $j(T, \mu; V)$ 
approaches the $V$-independent one, $j(T, \mu)$,
as $V \to \infty$, 
i.e., 
$j(T, \mu ;V) = j(T, \mu) + o(1)$.
%in the thermodynamic limit.]
At finite $T$, 
since the entropy density 
$s = - \partial j/\partial T = \Theta(1)$,
we have
\begin{equation}
2 V \beta \{ j(T/2, \mu ;V)-j(T, \mu ;V) \}
% \beta \{ j(1/2\beta, \mu;V)-j(1/\beta, \mu ;V) \}
\simeq V s(T, \mu)
% =  
% \Theta(s(T, \mu))
=
\Theta(V).
\label{delta_j}
\end{equation}
Therefore, 
$B_V^2 
\leq 
1/e^{V s(T, \mu)} 
=
1/e^{\Theta(V)}$. % \cite{magDV1}.
Inserting this result into inequality (\ref{Markov}),
we find that  
% it is seen that 
$
\langle \beta \mu V | \beta \mu V \rangle
\stackrel{P}{\to}
\Xi(\beta, \mu, V)
$,
% which is finite for finite $V$,
where
`$\stackrel{P}{\to}$' denotes convergence in probability.
Since 
$\Xi$ is finite for finite $V$,
$| \beta \mu V \rangle$ is well-defined.
This argument also shows that a single realization of 
$|\beta \mu V \ket$ %the gTPQ state 
gives $j$ by %, %the free energy density $f(T; V)$,
\begin{equation}
- V \beta j(T, \mu ;V)
=
% (1/V) %{1 \over V} 
\ln \langle \beta \mu V |\beta \mu V \rangle,
\label{ln<>=j}
\end{equation}
with exponentially small probability of error.
All genuine thermodynamic variables, such as entropy, 
can be calculated from $j$.

We then show that $|\beta \mu V \ket$ is a gTPQ state, i.e., 
$
{\langle}\hat{A}{\rangle}^{\rm TPQ}_{\beta \mu V}
\stackrel{P}{\to}
{\langle}\hat{A}{\rangle}^{\rm ens}_{\beta \mu V}
$
uniformly for every mechanical variable $\hat{A}$ 
as $V \to \infty$,
where
$
{\langle}\hat{A}{\rangle}^{\rm TPQ}_{\beta \mu V}
\equiv
{\langle}{\beta \mu V}|\hat{A}|{\beta \mu V}\rangle
/
{\langle}{\beta \mu V}|{\beta \mu V}\rangle
$.
To see this, 
we take
$
x =
\langle \hat{A} \rangle^{\rm TPQ}_{\beta \mu V}
$
and
$
y =
\langle \hat{A} \rangle^{\rm ens}_{\beta \mu V}
$
in inequality (\ref{Markov}), 
and evaluate
$
D_V(A)^2
\equiv
\overline{
( %\left(
{\langle}\hat{A}{\rangle}^{\rm TPQ}_{\beta \mu V}
		-{\langle}\hat{A}{\rangle}^{\rm ens}_{\beta \mu V}
)^2 %\right)^2
}
$.
Dropping smaller-order terms, we find 
\begin{equation}
D_V(A)^2
\leq
{
\langle (\Delta \hat{A})^2 \rangle^{\rm ens}_{2\beta \mu V}
+
(\langle A \rangle^{\rm ens}_{2\beta \mu V} 
- \langle A \rangle^{\rm ens}_{\beta \mu V} )^2
\over 
\exp [2V\beta \{ j(T/2, \mu ;V)-j(T, \mu ;V) \}]},
% \exp [2V\beta \{ j(1/2\beta, \mu ;V)-j(1/\beta, \mu ;V) \}]},
\label{Var<M>}
\end{equation}
where
${\langle}(\Delta \hat{A})^2{\rangle}^{\rm ens}_{\beta \mu V}
{\equiv}{\langle}(\hat{A}-\langle{A}\rangle^{\rm ens}_{\beta \mu V})^2 
\rangle^{\rm ens}_{\beta \mu V}$,
and so on.
The denominator of the r.h.s $=e^{\Theta(V)}$ 
from Eq.~(\ref{delta_j}),
whereas the numerator  
$\leq \Theta(V^{2m})$
from (\ref{bound_<A>}).
Hence,
$D_V(A)^2 \leq V^{2m}/e^{\Theta(V)}$, 
% \cite{magDV2},
which vanishes exponentially fast with increasing $V$, 
for every mechanical variable $\hat{A}$.
Therefore, 
$
\langle \hat{A} \rangle^{\rm TPQ}_{\beta \mu V}
\stackrel{P}{\to}
\langle \hat{A} \rangle^{\rm ens}_{\beta \mu V}
$
uniformly,
which 
shows that $|\beta \mu V \rangle$ is a gTPQ state.
% This means that 
A single realization of 
the gTPQ state gives equilibrium values of mechanical variables, 
with exponentially small probability of error, 
by $\langle \hat{A} \rangle^{\rm TPQ}_{\beta \mu V}$.

Note that one can use any convenient basis as $\{ | \nu \ket \}_\nu$,
because
the above construction of $|\beta \mu V \ket$ 
is independent of the choice of 
the basis. % $\{ | \nu \ket \}_\nu$,
Moreover, using $j$ obtained from formula (\ref{ln<>=j}),
one can estimate the upper bounds of errors 
from formulas (\ref{Var<>}) and (\ref{Var<M>}),
without resorting to results of other methods.
% In this sense, our formulas % new formulation % using the TPQ state 
% are almost self-validating ones. %a self-contained theory. 
This self-validating property 
is particularly useful in practical applications.

% In a way similar 
Similarly to the above construction of the gTPQ state, 
we can also generalize the cTPQ state proposed in Ref.~\cite{SS2013}
so as to be applicable to systems with $\dim \mathcal{H} = \infty$.

{\em Practical computational method -- }
% {\em Practical formulas -- }
The TPQ formulation sheds new light on quantum statistical mechanics
because it is much different from the ensemble formulation \cite{Kindai2013}.
For example, the von Neumann entropy, 
which coincides with the thermodynamic entropy
in the ensemble formulation, vanishes for TPQ states.
Because of this great difference, 
the TPQ formulation will also be useful for practical computations.
% That is, 
% since all other computational methods aim to compute formulas of 
% the ensemble formulation, 
% new possibilities are opened if 
% the formulas of the TPQ formulation are aimed for.
%%
% To show that this is indeed the case, 
To make this visible, 
we have developed
practical formulas that are particularly useful for 
numerical computations. 
They are presented in Ref.~\cite{SM}.
% Supplemental Material at [URL will be inserted by publisher].
Using them, %these practical formulas, 
one can obtain $|\beta \mu V \rangle$ simply 
by multiplying $[\mbox{constant} - (\hat{H} - \mu \hat{N})]$ 
% by multiplying $(l - \hat{g})$ 
with a random vector 
repeatedly $\Theta(N)$ times.
This is a powerful numerical method, as evidenced below.

{\em Application to the Hubbard model -- } 
% {\em Application to the Hubbard chain -- }
We now apply the present formulation to strongly-interacting electrons.
We take the Hubbard model
$ %\begin{equation}
\hat{H}
=
- \sum_{\langle \bm{r}, \bm{r}' \rangle} 
(\hat{c}_{\bm{r} \sigma}^\dagger \hat{c}_{\bm{r}' \sigma} + h.c.)
+ U \sum_{\bm{r}} 
(\hat{n}_{\bm{r} \uparrow} -1/2) (\hat{n}_{\bm{r} \downarrow} -1/2) 
% - \mu \hat{N}
$ %\end{equation}
with the periodic boundary conditions,
% which is % considered to be 
% obtained after cutting off a high energy part
% of certain electron systems.
where $\langle \bm{r}, \bm{r}' \rangle$ denotes a nearest pair of sites.
We consider a $1d$ chain and a $2d$ triangular lattice.
The number of sites $V$ is taken as 
$V = 14, 15$ because of the size of 
the memory of our computers.
Although this is larger than $V$ 
of the numerical diagonalization (ND) ever performed
(of the full spectrum to compute finite-temperature properties),
% approaches for finite temperature,
the factor of Eq.~(\ref{delta_j}), which appears in the r.h.s. of 
(\ref{Var<>}) and (\ref{Var<M>}), is not large enough.
In such a case, one can reduce errors by averaging 
the denominators and numerators, separately,
of these formulas over many realizations 
of the gTPQ states.
Averaging over $M$ realizations reduces the error
% as measured by the standard deviation,
by the factor of $1/\sqrt{M}$.
% We therefore take averages over $M$ realizations ($10 \leq M \leq 26$)
% to reduce errors, which is proportional to $1/\sqrt{M}$ as mentioned above.
[By contrast, averaging was not necessary for the spin system of Ref.~\cite{SS2013} because $V$ ($=27,30$) was large enough.]
We here take $10 \leq M \leq 26$. 

\begin{figure}[htbp]
\begin{center}
\includegraphics[width=0.9\linewidth]{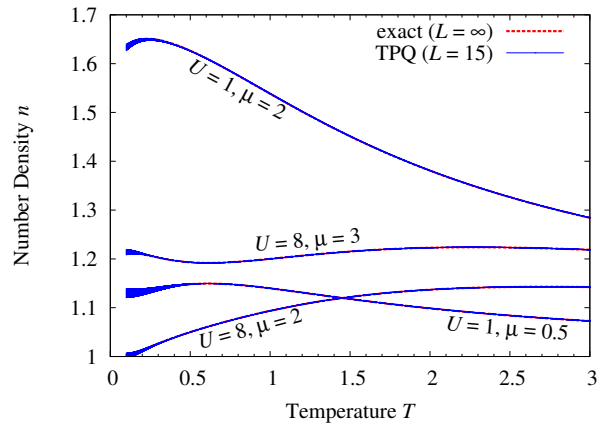}
\vspace{-6mm}
\end{center}
\caption{
$n$ versus $T$, obtained by the gTPQ states with $L=15$,
for $(U,\mu,M)=(8,2,18)$, $(8,3,20)$, $(1,0.5,18)$, and $(1,2,22)$.
Error bars show estimated errors, 
which can be made smaller by increasing $M$.
% Estimated errors, 
% which can be made smaller by increasing $M$, 
% are {\bf of the same order as the line width}.
% for 
% $U=8,\mu=2$ ($M=144$),
% $U=8,\mu=3$ ($M=116$),
% $U=1,\mu=0.5$ ($M=25$),
% $U=1,\mu=2$ ($M=32$).
Exact results for $L=\infty$ are also plotted.
}
\label{fig1}
\end{figure}

\begin{figure}[htbp]
\begin{center}
\includegraphics[width=0.9\linewidth]{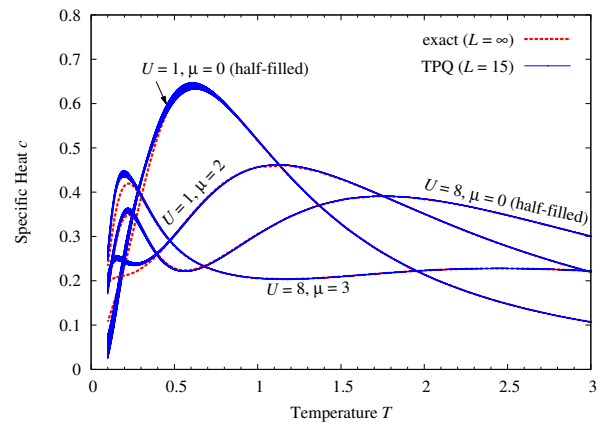}
\vspace{-6mm}
\end{center}
\caption{$c$ versus $T$,
obtained by the gTPQ states with $L=15$,
for $(U,\mu,M) = (8,0,14)$, $(8,3,20)$, $(1,0,12)$, and $(1,2,22)$.
Error bars show
estimated errors, 
which can be made smaller by increasing $M$.
Exact results for $L=\infty$ are also plotted.
}
\label{fig2}
\end{figure}

\begin{figure}[htbp]
\begin{center}
\includegraphics[width=0.9\linewidth]{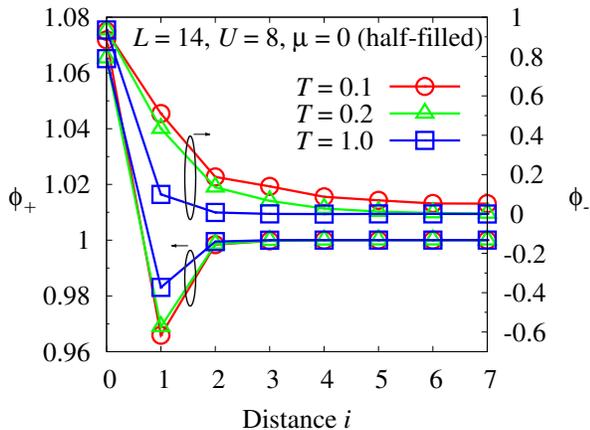}
\vspace{-6mm}
\end{center}
\caption{
$\phi_\pm$ versus $i$  for $U=8, \mu=0$,
obtained by the gTPQ states with $L=14$ and $M=21$,
at $T=0.1, 0.2, 1.0$.
% Exact results are unknown.
}
\label{fig3}
\end{figure}

\begin{figure}[htbp]
\begin{center}
\includegraphics[width=0.9\linewidth]{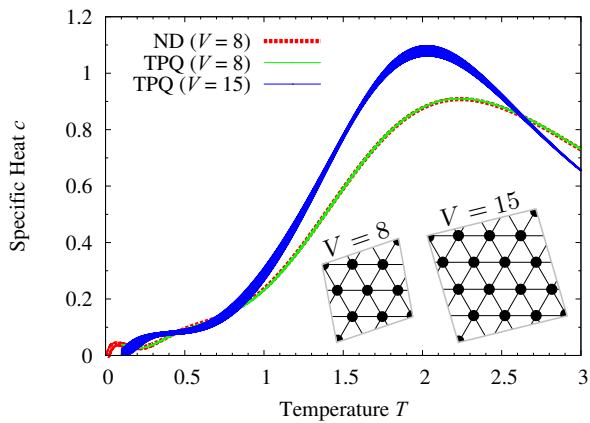}
% \setlength\unitlength{.9\linewidth}
% \begin{picture}(1,.7)(0,0)
% 	\put(0,0){\includegraphics[width=\unitlength]{fig4.eps}}
% 	\put(0.52,0.13){\includegraphics[width=.15\unitlength]{fig4-a.eps}}
% 	\put(0.52,0.27){\rotatebox{19.11}{$L=8$}}
% 	\put(0.69,0.13){\includegraphics[width=.23\unitlength]{fig4-b.eps}}
% 	\put(0.71,0.32){\rotatebox{13.90}{$L=15$}}
% \end{picture}
\vspace{-6mm}
\end{center}
\caption{$c$ versus $T$, for the $2d$ triangular lattice
with $U=3, \mu=1$,
% with $V=8$ and $15$,
% (shown in the inset) 
obtained by the gTPQ states with $V=8$ ($M=1024$) and with $V=15$ ($M=10$), 
and by ND for $V=8$.
% with 
% $(V,U,\mu,M) = (8,3,1,1024)$ and $(V,U,\mu,M) = (15,3,1,10)$.
Error bars show
estimated errors, 
which can be made smaller by increasing $M$.
% Here we take $M=1024$ for $V=8$, and $M=10$ for $V=15$.
}
\label{fig4}
\end{figure}

\begin{figure}[htbp]
\begin{center}
\includegraphics[width=0.9\linewidth]{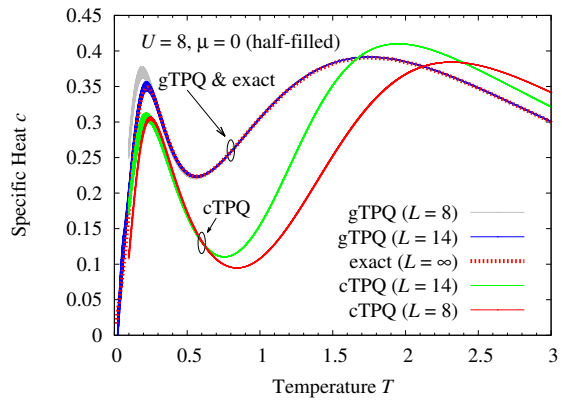}
\vspace{-6mm}
\end{center}
\caption{$c$ versus $T$, 
obtained by the gTPQ states with $L=8$ ($M=1024$) and $L=14$ ($M=26$), 
and by cTPQ states with $L=8$ ($M=1024$) and $L=14$ ($M=22$).
Error bars show estimated errors, 
which can be made smaller by increasing $M$.
Exact results for $L=\infty$ are also plotted.
}
\label{fig5}
\end{figure}

We first study the $1d$ chain of length $L$ ($=V$) 
% We first study the $1d$ system 
% (for which $V=L$, the length of the chain) 
as a benchmark, 
because some of physical quantities were exactly obtained 
for $L=\infty$ \cite{exact}.
% Here, the system size is denoted by $L$ instead of $V$.
Since the results for $U<0$ can be obtained from those for $U>0$
(see, e.g., Refs.~\cite{U-symmetry1,U-symmetry2}), 
% (see Supplementary Material ???),
we can assume $U>0$ without loss of generality.
We here take two values;
$U=1$, where the wave-particle duality plays essential roles,
and $U=8$, where the particle nature is stronger.
Regarding $\mu$, 
it can be controlled in experiments by an external voltage \cite{FET1,FET2,FET3}, in which $\mu$ is the electro-chemical potential.
Hence, we take several values;
$\mu=0$ (half-filled), $0.5, 2, 3$. % and $\mu=@@@$ (doped).
$T$ is taken as $0.1 \leq T \leq 3$ (Figs.~\ref{fig1}-\ref{fig4})
and $0.03 \leq T \leq 3$ ($L=14$ by the gTPQ state in Fig.~\ref{fig5}).
% $T$ is taken as $0.03$ or $0.1 \leq T \leq 3$.
To the authors' knowledge, 
no other numerical methods have ever succeeded in analyzing 
the Hubbard chain over such wide 
ranges of $T, \mu, U$ (see, e.g., Ref.~\cite{tohyama}).
One can go down to even lower $T$ by increasing 
the computational parameters $k_{\rm term}$
(defined in Ref.~\cite{SM}) and $M$.

The % results for the 
particle density
$n \equiv N/L$, obtained using the gTPQ states with $L=15$,    
is plotted in Fig.~\ref{fig1}.
[We take $\mu \neq 0$ because $\mu=0$ gives the trivial result $n=1$.]
The results agree well with 
% For comparison, we have also plotted 
the exact results for the infinite system $L \to \infty$ (broken lines)
\cite{exact}.
We also calculate 
the specific heat at constant $\mu$, 
defined by 
$c \equiv (T/L) (\partial S/\partial T)_{\mu,L}$.
% $c$ (at constant $N,L$) 
Generally, $c$ is much harder to compute than $n$
because $c$ is a higher (second) derivative of $j$.
% because they are a second and first derivatives, respectively, of $j$.
As shown in Fig.~\ref{fig2}, 
the results of the gTPQ states with $L=15$ agree fairly well with 
the exact results for $L \to \infty$ (broken lines) \cite{exact}.
Small deviations are due to finite-size effects, 
as will be discussed later.

Furthermore, we calculate correlation functions, 
for which exact solutions are unknown.
We calculate the charge and 
the staggered spin correlation functions
$\phi_+$ and $\phi_-$,
% $\phi_+$ (Fig.~\ref{fig3}) and $\phi_-$ (Fig.~\ref{fig5}),
% the spin, charge and spin-charge correlation functions,
respectively, 
% which are 
which are defined by
\begin{equation}
\phi_{\pm}(i)
\equiv
{(\pm 1)^i \over L}
\!
\sum_{j} 
\bra 
(\hat{n}_{j \uparrow} \pm \hat{n}_{j \downarrow})
(\hat{n}_{j+i \uparrow} \pm \hat{n}_{j+i \downarrow})
\rangle_{\beta \mu L}.
\end{equation}
As shown in Fig.~\ref{fig3}, 
$\phi_+$ %in Fig.~\ref{fig3} 
has a dip at $i=1$, whereas
$\phi_-$ %in Fig.~\ref{fig5} 
decreases monotonically with increasing $i$.
These behaviors are manifestations of 
% can be understood as a result of 
the wave-particle duality. 
$\phi_-$ was previously computed numerically in Ref.~\cite{tohyama},
where $T$ was limited to $T \leq 0.2$ and $\phi_+$ was not computed.
Our results agree well with theirs.

We then study the $2d$ triangular lattice, for which exact results are unknown.
We analyze a weakly doped case ($0 < \mu \ll$ band width),
which will be most interesting experimentally,
over a wide range of $T$.
Such a case is hard to analyze with most numerical methods
because of the sign problem and so on.
We first solve a small system with $V=8$,
for which ND of the full spectrum is possible.
In Fig.~\ref{fig4}, the results for the specific heat $c$, 
obtained with ND and the gTPQ states,
are plotted as a function of $T$.
The agreement is very good.
We then solve a larger system with $V=15$,
for which ND of the full spectrum is impossible.
The result obtained with the gTPQ states is plotted in Fig.~\ref{fig4}.
Since we have rigorously proved that the gTPQ states give 
correct results (for each finite $V$) with high probability, 
our result is reliable within the error bars, which 
can be made arbitrarily small by increasing $M$ (the number of realizations).
That is, we have successfully obtained 
% We therefore believe that 
reliable results for $V=15$ 
over a wide range of $T$.
% is obtained. % here for the first time.

{\em Superiority of the gTPQ state -- } % and grand-canonical ensemble-- } 
% {\em Advantages of gTPQ state -- } 
% {\em Discussions -- } 
We have rigorously proved that 
the results of a TPQ state of size $V$ 
agree with those of the corresponding Gibbs state
{\em of the same size $V$},
within exponentially small error. 
% On the other hand, % the finite-size effect, i.e., 
However, generally, these results for a finite-size system 
deviate from those for an infinite system.
Typically, this finite-size effect 
%such deviation %, i.e., the finite-size effect, 
is inversely proportional to a power of $V$,
and hence is not so small in general.
Then a question arises: 
Which TPQ state has a smaller finite-size effect,
the gTPQ state or the cTPQ state?

To answer this question,
we compute $c$ of the $1d$ chain % at $\mu=0$ (half-filled)
for $L=8$ and $14$, using both TPQ states.
We take $\mu=0$ (half-filled), for which 
$n$ is independent of $T$ ($n=1$) and hence 
$c$(at constant $\mu$) $=$ $c$(at constant $N$) for $L = \infty$.
The results are plotted in Fig.~\ref{fig5}. 
We find that 
% As seen from Fig.~\ref{fig5}, 
the finite-size effect is much smaller in the gTPQ states than 
in the cTPQ states.
Even for $L=8$, the result of the gTPQ state is surprisingly close to 
the exact result for $L=\infty$.
By contrast, 
% deviation of the result of the cTPQ state is very large.
% deviate significantly from the exact results for $L=\infty$.
% [For $L=8$, we have confirmed using ND that  
% these results of the gTPQ and cTPQ states agree well with 
% those of the grand-canonical and canonical Gibbs states, 
% respectively.] %, within the error bars.]
% This 
the cTPQ states have very large finite-size effects even for $L=14$.
That is, %We thus find that 
the gTPQ state simulates a finite subsystem in an infinite system
much better than the cTPQ state.
This seems reasonable because
the gTPQ state contains information about all values of $N$
whereas the cTPQ state contains information 
only about a specific value of $N$.
% In addition to smaller finite-size effects, 
Moreover, the gTPQ state also has another advantage that one can study an 
arbitrary value of the filling factor $N/2L$ 
% $\bra N \ket / 2L$ 
for any $L$.
For example, one can calculate the quarter-filled case,
where $N=L/2$, 
 % $N/2L = 1/4$ 
even for odd $L$.
This makes wider the available ranges of parameters in numerical 
computations. % than the cTPQ state.
For these reasons, the gTPQ state 
would be far superior for practical purposes.
If one has to use the cTPQ state (e.g., to save computer 
resources), 
it is better to convert the results % to those of the gTPQ state 
using, for example, % Eq.~(\ref{conv_c2g}).
the relation %fact that 
$
\sum_N \langle \beta N V | \beta N V \rangle e^{\beta \mu N}
\stackrel{P}{\to}
\Xi(\beta, \mu, V).
$

These conclusions also apply to 
comparison between the canonical and grand-canonical 
Gibbs states in the ensemble formulation, 
because their results are identical to those of the cTPQ and gTPQ states, respectively (with exponentially small errors).
To the authors' knowledge, systematic studies on such comparison 
were not reported previously
% This is probably 
because $V$ of ND of the full spectra
is severely upper bounded.

% Finally, note that for the $L=14$ gTPQ state 
% in Fig.~\ref{fig5} we have taken data for $T$
% down to $0.03$ in order to demonstrate that the TPQ formulation 
% gives good results even at such low temperature.
% One can go down to even lower $T$ by increasing 
% the computational parameters $k_{\rm term}$
% (defined in Ref.~\cite{SM})
% and $M$.

% Finally, we stress that the TPQ formulation is applicable to any physical systems (if they are consistent with thermodynamics).
% Hence, we expect that it will solve many problems that are 
% hard to solve with other methods, 
% as already demonstrated in Ref.~\cite{SS2013} for a spin system.

\begin{acknowledgments}
We thank % are grateful to 
C. Hotta and H. Tasaki for helpful discussions.
This work was supported by KAKENHI Nos.~22540407, 24540393 and 26287085.
SS is supported by JSPS Research Fellowship No.~245328.
Although all the numerical results in this Rapid Communication have been obtained 
using workstations, 
preliminary computation has been done using the facilities of the Supercomputer Center, the Institute for Solid State Physics, the University of Tokyo.
\end{acknowledgments}

% \bibliography{basename of .bib file}

\begin{thebibliography}{99}
\bibitem{degreeA}
Actually, our formula for the probabilistic errors show
that the TPQ states give the correct results 
for more general $\hat{A}$,
such as
$| \langle{ \hat{A}^2 }\rangle^{\rm ens}_{\beta \mu V} |
\leq 
Ke^{o(V)}
$
and 
$\hat{A} = e^{i \xi \hat{a}}$,
where $\xi \in \mathbb{R}$ and $\hat{a}$ is an arbitrary observable.
\end{thebibliography}

\end{document}